\documentclass[pra,reprint,twocolumn,showpacs]{revtex4-1}

\usepackage{graphicx}

\usepackage{amsmath}

\usepackage{amssymb}

\usepackage[dvipsnames]{color}

\newcommand{\mbf}[1]{\mathbf{#1}}
\newcommand{\norm}[1]{\left|\left|#1\right|\right|}
\newcommand{\abs}[1]{\left|#1\right|}

\begin{document}

%\preprint{}

\title{Metric Space Formulation of Quantum Mechanical Conservation Laws}


\author{P. M. Sharp and I. D'Amico}
\email[corresponding author: irene.damico@york.ac.uk]{}
%\thanks{}
\affiliation{Department of Physics, University of York, York, YO10 5DD, United Kingdom}


\date{\today}

\begin{abstract}
We show that conservation laws in quantum mechanics naturally lead to metric spaces for the set of related physical quantities. All such metric
spaces have an ``onion-shell'' geometry. We demonstrate the power of this approach by considering many-body systems immersed in a magnetic field,
with a finite ground state current. In the associated metric spaces we find regions of allowed and forbidden distances, a ``band structure'' in
metric space directly arising from the conservation of the $z$ component of the angular momentum.
\end{abstract}

%\keywords{}

\maketitle


\section{Introduction}

Conservation laws are a central tenet of our understanding of the physical world. Their tight relationship to natural symmetries was demonstrated by
Noether in 1918 \cite{Noether1918} and has since been a fundamental tool for developing theoretical physics. In this paper we demonstrate how
these laws induce appropriate ``natural'' metrics on the related physical quantities. Conservation laws are central to the behavior of physical
systems and we show how this relevant physics is translated into the metric analysis. We argue that this alternative picture provides a new powerful
tool to study certain properties of many-body systems, which are often complex and hardly tractable when considered within the usual coordinate
space-based analysis, while may become much simpler when analyzed within metric spaces. We exemplify this concept by considering functional
relationships fundamental to current density functional theory (CDFT) \cite{Vignale1987,Vignale1988}.

We will first introduce a way to derive appropriate ``natural'' metrics from a system's conservation laws. Second, as an example application of the
approach, we will explicitly consider an important class of systems -- systems with applied external magnetic fields. In contrast with those to
which standard density functional theory (DFT) \cite{Dreizler1990} can be applied, systems subject to external magnetic fields are not simply
characterized by their particle densities as even their ground states may display a finite current \cite{Vignale1987,Vignale1988}. These systems are
of great importance, e.g., due to the emerging quantum technologies of spintronics and quantum information where, for example, few electrons in
nano- or microstructures immersed in magnetic fields are proposed as hardware units
\cite{Takahashi2010,daSilva2009,Brandner2013,Amaha2013,Castellanos-Beltran2013}.

To analyze systems immersed in a magnetic field, we will introduce a metric associated with the paramagnetic current density, which can be
associated with the angular momentum components. We will show that, at least for systems which preserve the $z$ component of the angular momentum,
the paramagnetic current density metric space displays an ``onion-shell'' geometry, directly descending from the related conservation law. In recent
work \cite{D'Amico2011,Artacho2011,D'Amico2011b} appropriate metrics for characterizing wavefunctions and particle densities within quantum
mechanics were introduced. It was shown that wavefunctions and their particle densities both form metric spaces with an ``onion-shell''
structure \cite{D'Amico2011}. We will show that, within the same general procedure used for the paramagnetic current, these metrics descend from the
respective conservation laws. We will then focus on ground states and characterize them not only through the mapping between wavefunctions and
particle densities, but importantly through mappings involving the paramagnetic current density. In fact, for systems with an applied magnetic field,
ground state wavefunctions are characterized uniquely only by knowledge of both particle \emph{and} paramagnetic current densities (and vice versa),
as demonstrated within CDFT \cite{Vignale1987,Vignale1988}.

The rest of this paper is organized as follows: In Sec. \ref{metric} we introduce our general approach to derive metric spaces from conservation
laws. We demonstrate the application of this approach to wavefunctions, particle densities, and paramagnetic current densities in Sec. \ref{apply}.
We consider systems subject to magnetic fields in Sec. \ref{cdft}. Here we use the metrics derived from our approach to study the fundamental
theorem of CDFT. We present our conclusions in Sec. \ref{conclusion}.

\section{Derivation of Metric Spaces from Conservation Laws} \label{metric}

A metric or distance function $D$ over a set $X$ satisfies the following axioms for all $x,y,z \in X$ \cite{Megginson1998,Sutherland2009}:
\begin{align}
 D(x,y) &\geqslant 0\ \text{and}\ D(x,y)=0 \iff x=y, \label{axiom1}\\
 D(x,y) &= D(y,x), \label{axiom2}\\
 D(x,y) &\leqslant D(x,z)+D(z,y), \label{axiom3}
\end{align}
with (\ref{axiom3}) known as the triangle inequality. The set $X$ with the metric $D$ forms the metric space $(X,D)$. It can be seen from the axioms
(\ref{axiom1}) - (\ref{axiom3}) that many metrics could be devised for the same set, some trivial. Here we introduce ``natural'' metrics associated
to conservation laws: this will avoid arbitrariness and in turn will ensure that the proposed metrics stem from core characteristics of the systems
analyzed and contain the related physics.

In quantum mechanics, many conservation laws take the form
\begin{equation} \label{conservation}
 \int \abs{f(x)}^{p} dx = c
\end{equation}
for $0<c<\infty$. For each value of $1\leqslant p<\infty$, the entire set of functions that satisfy (\ref{conservation}) belong to the $L^p$ vector
space, where the standard norm is the $p$ norm \cite{Megginson1998}
\begin{equation} \label{lp_norm}
 \norm{f(x)}_p =\left[\int \abs{f(x)}^{p} dx \right]^{\frac{1}{p}}.
\end{equation}
From any norm a metric can be introduced in a standard way as $D(x,y)=\norm{x-y}$ so that with $p$ norms we get
\begin{equation} \label{lp_metric}
 D_{f}(f_1,f_2):=\norm{f_1-f_2}_p.
\end{equation}
However before assuming this metric for the physical functions related to the conservation laws, an important consideration must be made: Eq.
(\ref{lp_metric}) has been derived assuming the ensemble $\{f\}$ to be a vector space; this is in fact necessary to introduce a norm. If we want to
retain the metric (\ref{lp_metric}), but restrict it to the ensemble of \emph{physical} functions satisfying (\ref{conservation}), which does not
necessarily form a vector space, we must show that (\ref{lp_metric}) is a metric for this restricted function set. This can be done using the general
theory of metric spaces: given a metric space $(X,D)$ and $S$ a non empty subset of $X$, $(S,D)$ is itself a metric space with the metric $D$
inherited from $(X,D)$. The metric axioms (\ref{axiom1}) - (\ref{axiom3}) automatically hold for $(S,D)$ because they hold for $(X,D)$
\cite{Megginson1998,Sutherland2009}. Hence, we have a metric for the functions of interest, as their sets are non empty subsets of the respective
$L^p$ sets.

The metric (\ref{lp_metric}) is then the one that \emph{directly descends} from the conservation law (\ref{conservation}). Conversely any
conservation law which can be recast as (\ref{conservation}) (for example conservation of quantum numbers) can be interpreted as inducing a
metric on the appropriate, physically relevant, subset of $L^{p}$ functions. This provides a general procedure to derive ``natural'' metrics from
physical conservation laws.

\section{Applications of the Metric Space Approach} \label{apply}

We now consider specific quantum mechanical functions and conservation laws. Following Ref. \cite{D'Amico2011} we use a convention where
wavefunctions are normalized to the particle number $N$ \footnote{This allows the description of Fock space as a set of concentric spheres}. Then
the particle density of an $N$-particle system and its paramagnetic current density are defined as
\begin{align}
 \rho(\mbf{r})&=\int \abs{\psi\left(\mbf{r},\mbf{r}_{2},\ldots,\mbf{r}_{N}\right)}^{2} d\mbf{r}_{2}\ldots d\mbf{r}_{N},\label{density}\\
 \mbf{j}_{p}(\mbf{r})&=-\frac{i}{2}\int \left(\psi^{\ast}\nabla\psi - \psi\nabla\psi^{\ast}\right) d\mbf{r}_{2}\ldots d\mbf{r}_{N}.\label{current}
\end{align}
First of all we note that $\psi\left(\mbf{r}_1,\mbf{r}_{2},\ldots ,\mbf{r}_{N}\right)$ and $\rho(\mbf{r})$ are subject to the following conservation laws
(wavefunction norm and particle conservation):
\begin{align}
 &\int\abs{\frac{\psi\left(\mbf{r}_1,\mbf{r}_{2},\ldots ,\mbf{r}_{N}\right)}{\sqrt{N}}}^{2}d\mbf{r}_{1}\ldots d\mbf{r}_{N} = 1,\label{psi_cons}\\
 &\int\rho(\mbf{r}) d\mbf{r} = N.\label{rho_cons}
\end{align}
Similarly the paramagnetic current density $\mbf{j}_{p}(\mbf{r})$ obeys
\begin{equation} \label{Lz}
 \int \left[\mbf{r}\times\mbf{j}_{p}(\mbf{r})\right]_z d\mbf{r} = \langle\psi|\hat{L}_z|\psi\rangle.
\end{equation}
For eigenstates of systems for which the $z$ component of the angular momentum is preserved we then have $\langle\hat{L}_z\rangle=m$, with $m$ an
integer, and (\ref{Lz}) can be recast as
\begin{equation} \label{j_p_cons}
 \int \abs{\left[\mbf{r}\times\mbf{j}_{p}(\mbf{r})\right]_z} d\mbf{r} = \abs{m}.
\end{equation}
For wavefunctions and particle densities our procedure leads to the metrics introduced in Ref. \cite{D'Amico2011} ($N$ fixed)
\cite{Artacho2011,D'Amico2011b}
\begin{align}
 D_{\psi}(\psi_{1},\psi_{2})=&\left[\int \left(\abs{\psi_{1}}^{2}+\abs{\psi_{2}}^{2}\right)d\mbf{r}_1\ldots d\mbf{r}_{N}\right.\nonumber\\
 &- \left. 2\abs{\int\psi_{1}^{*}\psi_{2}d\mbf{r}_1\ldots d\mbf{r}_{N}}\right]^{\frac{1}{2}},\label{dpsi}\\
 D_{\rho}(\rho_{1},\rho_{2})=&\int\abs{\rho_{1}(\mbf{r})-\rho_{2}(\mbf{r})} d\mbf{r}; \label{drho}
\end{align}
for the paramagnetic current density, our procedure introduces the following metric:
\begin{equation} \label{dj_p}
 D_{\mbf{j}_{p_{\perp}}}(\mbf{j}_{p,1},\mbf{j}_{p,2})=\int\abs{\left\{\mbf{r}\times\left[\mbf{j}_{p,1}(\mbf{r})-\mbf{j}_{p,2}(\mbf{r})\right]\right\}_{z}} d\mbf{r}.
\end{equation}
We note that $D_{\mbf{j}_{p_{\perp}}}$ will be a distance between equivalence classes of paramagnetic currents, each class characterized by
current densities having the same transverse component $\mbf{j}_{p_{\perp}}\equiv(j_{p,x},j_{p,y})$. $D_{\mbf{j}_{p_{\perp}}}$ is gauge
invariant provided that $\mbf{j}_{p,1}$ and $\mbf{j}_{p,2}$ are within the same gauge and $[\hat{L_{z}},\hat{H}]=0$.

Next we show that conservation laws naturally build within the related metric spaces a hierarchy of concentric spheres, or ``onion-shell'' geometry.
If we set as the center of each sphere the zero function $f^{(0)}(x)\equiv 0$, and consider the distance between it and any other element in the
metric space, we recover the $p$-norm expressions (\ref{lp_norm}) directly descending from the related conservation laws. This procedure induces in
the related metric spaces a structure of concentric spheres with radii, in the cases considered here, of natural numbers to the power of $1/p$: all
functions corresponding to the same value of a certain conserved quantity will lay on the surface of the same sphere. Specifically, for systems of
$N$ particles, wavefunctions lie on spheres of radius $\sqrt{N}$, and particle densities on spheres of radius $N$; for the metric space of
paramagnetic current densities, all paramagnetic current densities with a $z$ component of the angular momentum equal to $\pm m$ lie on spheres of
radius $\abs{m}$.

The first axiom of a metric (\ref{axiom1}) guarantees that the minimum value for all distances is $0$, and that this value is attained for two
identical states. The onion-shell geometry guarantees that, for functions on the surface of the same sphere, i.e., which satisfy a certain
conservation law with the same value, there is also an upper limit for their distance associated with the diameter of the sphere. From (\ref{dj_p})
we see that for paramagnetic current densities this upper limit is achieved in the limit of currents which do not spatially overlap. This is also
the case for particle densities, as seen in (\ref{drho}).
\begin{figure}[t]
 \includegraphics[width=\columnwidth]{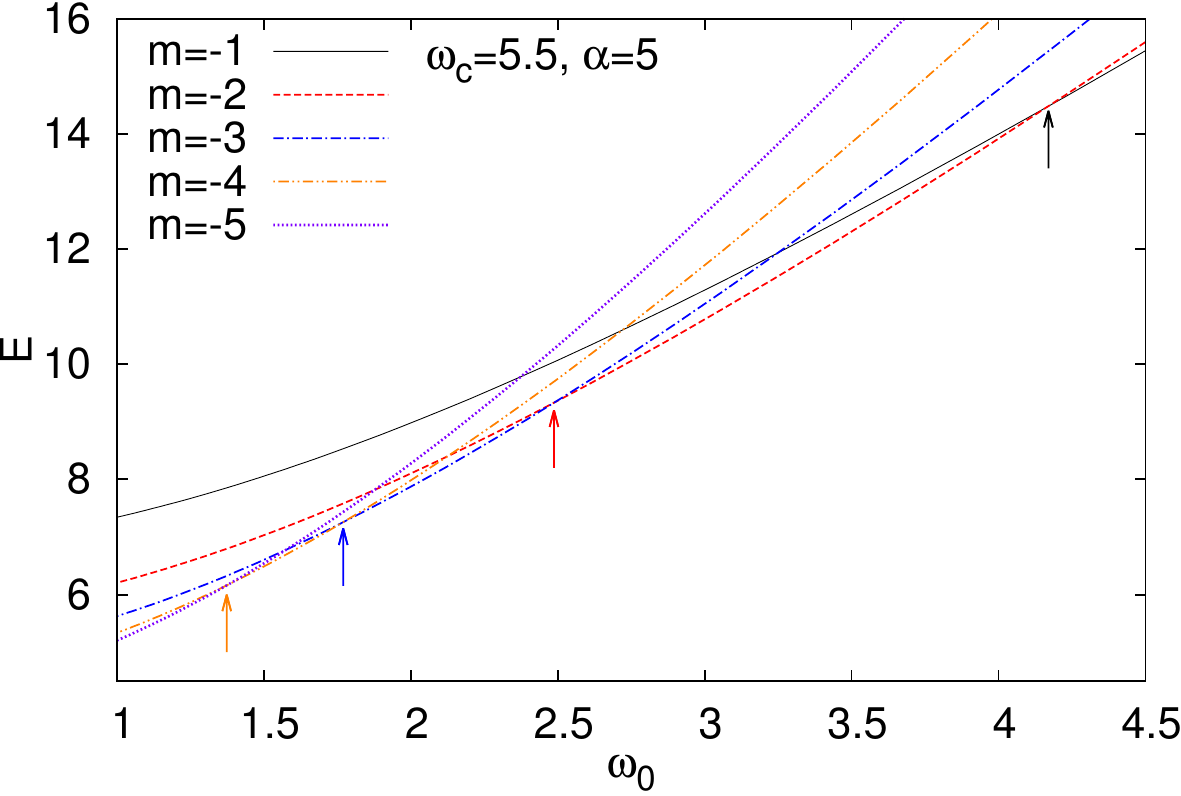}
 \caption{(Color online) For the ISI system energy is plotted against the confinement frequency for several values of the angular momentum quantum number $m$ (as
 labeled), and with constant cyclotron frequency and interaction strength. Arrows indicate where the value of $m$ for the ground state changes.}
 \label{energy}
\end{figure}

Interestingly, and in contrast to wavefunctions and particle densities \cite{D'Amico2011}, even when considering systems with the same number of
particles it may be necessary to consider paramagnetic current densities with different values of $m$; in terms of their metric space geometry,
current densities that have different values of $\abs{m}$ lie on different spheres. Therefore, the maximum value for the distance between
paramagnetic current densities of a system of $N$ particles is related to the upper limit of the number of spheres in the onion-shell geometry.
Using the triangle inequality we have in fact
\begin{align}
D_{\mbf{j}_{p_{\perp}}}(\mbf{j}_{p,m_{1}},\mbf{j}_{p,m_{2}})&\leqslant D_{\mbf{j}_{p_{\perp}}}(\mbf{j}_{p,m_{1}},\mbf{j}^{(0)}_{p})+D_{\mbf{j}_{p_{\perp}}}(\mbf{j}^{(0)}_{p},\mbf{j}_{p,m_{2}})\nonumber\\
&=\abs{m_1}+\abs{m_2}\leqslant l_{1}+l_{2},
\end{align}
where $l_i$ is the quantum number related to the total angular momentum of system $i$.

\section{Study of Model Systems} \label{cdft}

We now concentrate on the sets of ground state wavefunctions, related particle densities, and related paramagnetic
current densities. Since ground states are non empty subsets of all states, ground-state-related functions form metric spaces with the metrics
(\ref{dpsi}), (\ref{drho}), and (\ref{dj_p}). The importance of characterizing ground states and their properties has been highlighted by the huge
success of DFT (in all its flavors) as a method to predict devices' and material properties \cite{Dreizler1990,Ullrich2013}. Standard DFT is built
on the Hohenberg-Kohn (DFT-HK) theorem \cite{HK1964}, which demonstrates a one-to-one mapping between ground state wavefunctions and their particle
densities. This theorem is highly complex and nonlinear in coordinate space. However, Ref. \cite{D'Amico2011} showed that the DFT-HK theorem
is a mapping between metric spaces, and may be very simple when described in these terms, becoming monotonic and almost linear for a wide range of
parameters and for the systems there analyzed. CDFT is a formulation of DFT for systems in the presence of an external magnetic field. In
CDFT \cite{Vignale1987,Vignale1988} the original HK mapping is extended (CDFT-HK theorem) to demonstrate that $\psi$ is uniquely determined only by
knowledge of both $\rho(\mbf{r})$ and $\mbf{j}_p(\mbf{r})$ (and vice versa). This is the theorem we will consider in this section.

To further our analysis, we now explicitly examine two model systems with applied magnetic fields. They both consist of two electrons parabolically
confined that interact via different potentials, Coulomb (magnetic Hooke's atom) \cite{Taut2009} and inverse square interaction (ISI)
\cite{Quiroga1993}, respectively. Both systems may be used to model electrons confined in quantum dots. The Hamiltonians for the magnetic Hooke's
atom and the ISI system are
\begin{align}
 \hat{H}_{HA}&=\sum_{i=1}^{2}\left\{\frac{1}{2}\left[\hat{\mbf{p}}_{i}+\mbf{A}\left(\mbf{r}_{i}\right)\right]^{2}+\frac{1}{2}\omega_{0}^{2}r_{i}^{2}\right\}+\frac{1}{\abs{\mbf{r}_{2}-\mbf{r}_{1}}}, \label{Hooke_H}\\
 \hat{H}_{ISI}&=\sum_{i=1}^{2}\left\{\frac{1}{2}\left[\hat{\mbf{p}}_{i}+\mbf{A}\left(\mbf{r}_{i}\right)\right]^{2}+\frac{1}{2}\omega_{0}^{2}r_{i}^{2}\right\}+\frac{\alpha}{\left(\mbf{r}_{1}-\mbf{r}_{2}\right)^{2}}, \label{ISI_H}
\end{align}
(atomic units, $\hbar=m_e=e=1$). Here $\alpha$ is a positive constant, $\mbf{A}=\frac{1}{2}\mbf{B}\times\mbf{r}$ (symmetric gauge), and
$\mbf{B}=\omega_{c}c\mbf{\hat{z}}$ is a homogeneous, time-independent external magnetic field. For these systems $\langle\hat{L}_z\rangle$ is a
conserved quantity. Following Refs. \cite{Vignale1987,Taut2009} we disregard spin to concentrate on the features of the orbital currents. For
Hooke's atom, we obtain highly precise numerical solutions following the method in Ref. \cite{Coe2008}. The ISI system is solved exactly
\cite{Quiroga1993}.
\begin{figure*}[t]
 \includegraphics[width=\textwidth]{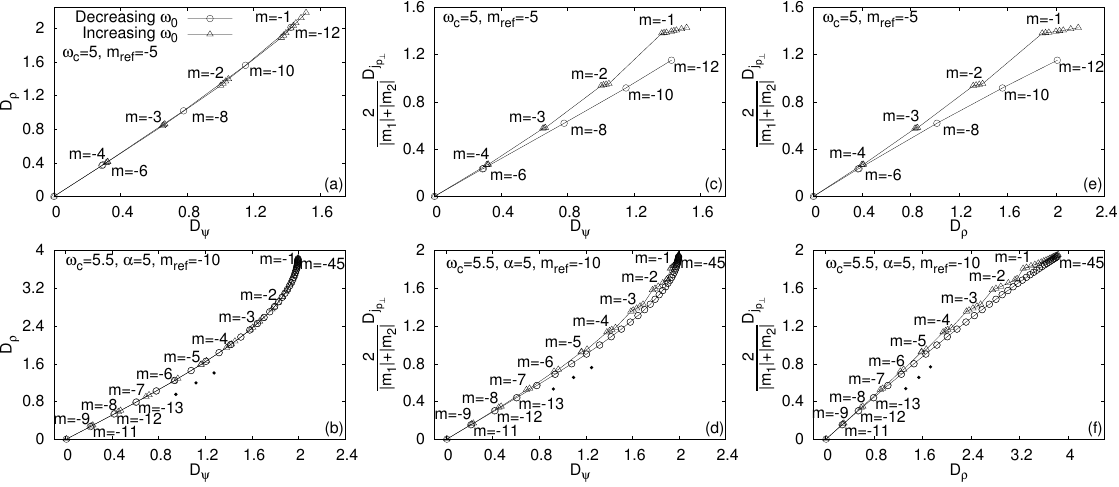}
 \caption{Results for ground states. Top: Hooke's atom (reference state ${\omega}_0=0.5, {\omega}_c=5, m_{ref}=-5$). Bottom: ISI system
 (reference state ${\omega}_0=0.62, {\omega}_c=5.5, {\alpha}=5, m_{ref}=-10$). Panels (a) and (b): $D_{\rho}$ vs $D_{\psi}$; (c) and (d): rescaled
 $D_{\mbf{j}_{p_{\perp}}}$ vs $D_{\psi}$; (e) and (f): rescaled $D_{\mbf{j}_{p_{\perp}}}$ vs $D_{\rho}$. Frequencies smaller than the reference
 are labeled with circles, larger with triangles.}
 \label{results}
\end{figure*}

\begin{figure}[t]
 \includegraphics[width=\columnwidth]{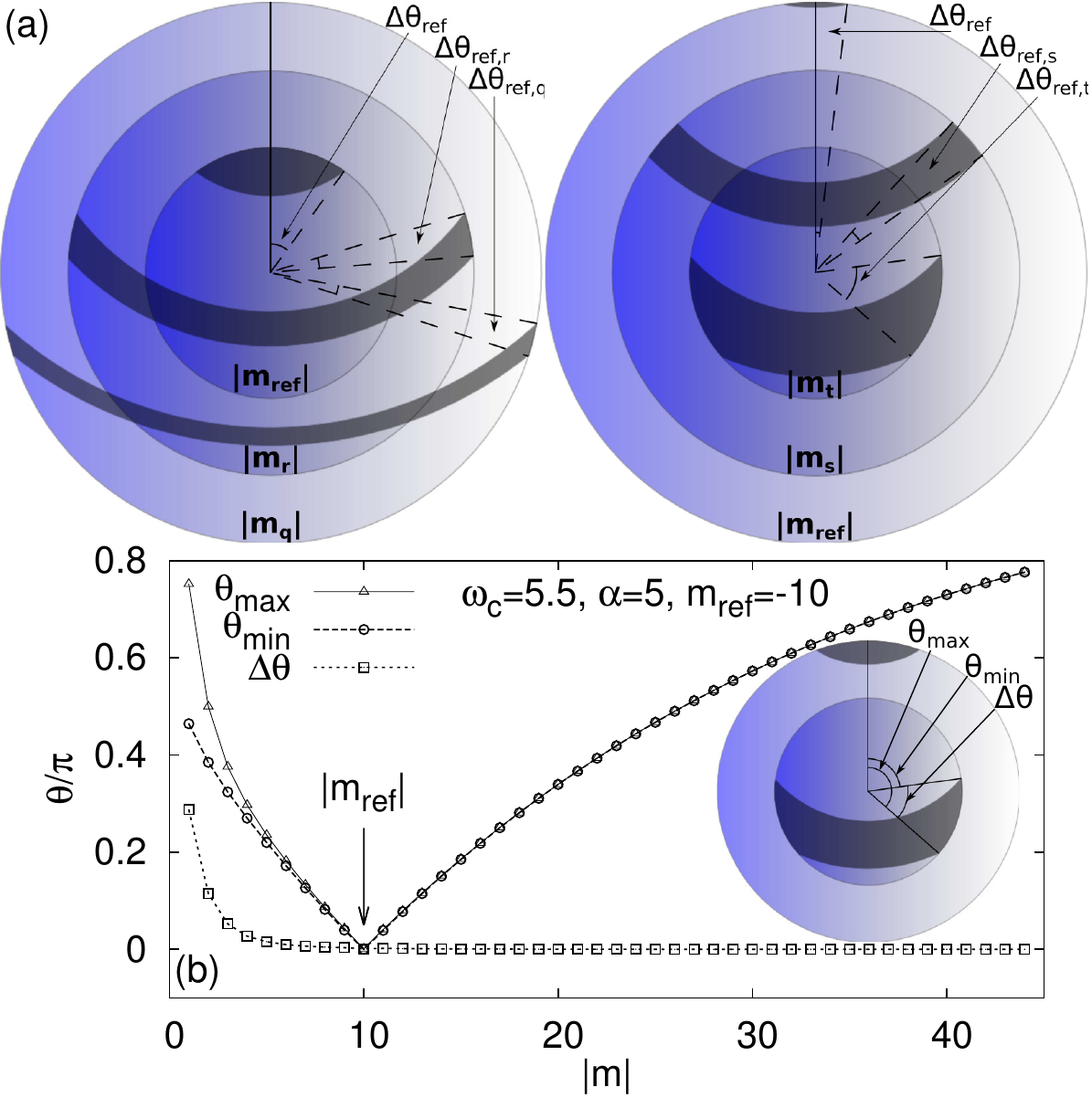}
 \caption{(Color online) (a) Sketch of the onion-shell geometry of the metric space for paramagnetic current densities, where
 $\abs{m_q}>\abs{m_r}>\abs{m_{ref}}$ (left) and $\abs{m_{ref}}>\abs{m_s}>\abs{m_t}$ (right). The reference state is at the north pole on the
 reference sphere. The dark gray areas denote the regions where ground state currents are located (`bands'), with dashed lines indicating their
 widths. (b) Results of the angular displacement of ground state currents for the ISI system. Lines are a guide to the eye. Inset: Definition of
 relevant angles.}
 \label{spheres}
\end{figure}

To produce families of ground states, for each system we systematically vary the value of $\omega_0$ (while keeping all other parameters constant),
and for each value we calculate the ground state wavefunction, particle density, and paramagnetic current density. A reference state is determined
by choosing a specific $\omega_0$ value, and the appropriate metric is then used to calculate the distances between it and each member of the
family. To ensure that we select ground states, varying $\omega_0$ may require varying the quantum number $m$ \cite{Taut2009,Quiroga1993}. This is
shown for the ISI system in Fig. \ref{energy}. Here, as $\omega_{0}$ increases, we must decrease the value of $\abs{m}$ in order to remain in the
ground state. As a result of this property, within each family of ground states, paramagnetic current densities will ``jump'' from one sphere of the
onion-shell geometry to another [see Fig.~\ref{spheres}(a), where the reference state is the `north pole' of its sphere]. To obtain ground states
with nonzero paramagnetic currents, we must use $\omega_0$ values corresponding to $m<0$  \cite{Taut2009,Quiroga1993}.

In Fig.~\ref{results}, we plot each pair of distances for the two systems. The reference states have been chosen so that most of the available
distance range can be explored both for the case of increasing and for the case of decreasing values of $\omega_0$. When considering the
relationship between ground state wavefunctions and related particle densities, Figs.~\ref{results}(a) and \ref{results}(b), our results confirm the
findings in Ref. \cite{D'Amico2011}: a monotonic mapping, linear for low to intermediate distances, and where vicinities are mapped onto vicinities;
also curves for increasing and decreasing $\omega_0$ collapse onto each other. However closer inspection reveals a fundamental difference with Ref.
\cite{D'Amico2011}, the presence of a ``band structure.'' By this we mean regions of allowed (``bands'') and forbidden (``gaps'') distances,
whose widths depend, for the systems considered here, on the value of $\abs{m}$. This structure is due to the changes in the value of the quantum
number $m$, which result in a substantial modification of the ground state wavefunction (and therefore density) and a subsequent large increase in
the related distances.

When we focus on the plots of paramagnetic current densities' against wavefunctions' distances, Figs.~\ref{results}(c) and \ref{results}(d), we find
that the ``band structure'' dominates the behavior. Here the change in $\abs{m}$ has an even stronger effect, in that
$dD_{\mbf{j}_{p_{\perp}}}/dD_{\psi}$ is noticeably discontinuous when moving from one sphere to the next in $\mbf{j}_p$ metric space. This
discontinuity is more pronounced for the path $\abs{m}<\abs{m_{ref}}$ than for the path $\abs{m}>\abs{m_{ref}}$. Similarly to Figs.~\ref{results}(a)
and \ref{results}(b), the mapping of $D_{\psi}$ onto $D_{\mbf{j}_{p_{\perp}}}$ maps vicinities onto vicinities and remains monotonic, but for small
and intermediate distances it is only piecewise linear. In contrast with $D_{\rho}$ vs $D_{\psi}$, curves corresponding to increasing and decreasing
$\omega_0$ do not collapse onto each other.

Figures~\ref{results}(e) and \ref{results}(f) show the mapping between particle and paramagnetic current density distances: this has characteristics
similar to the one between $D_{\psi}$ and $D_{\mbf{j}_{p_{\perp}}}$, but remains piecewise linear even at large distances.

We will now concentrate on the $\mbf{j}_{p}$ metric space to characterize the ``band structure'' observed in Fig.~\ref{results}. Within the metric
space geometry, we consider the polar angle $\theta$ between the reference $\mbf{j}_{p,ref}$ and the paramagnetic current density $\mbf{j}_{p}$ of
angular momentum $\abs{m}$. Using the law of cosines, $\theta$ is given by
\begin{equation} \label{angle}
 \cos{\theta}=\frac{m_{ref}^2+m^2-D_{\mbf{j}_{p_{\perp}}}^2(\mbf{j}_{p,ref},\mbf{j}_{p}) }{2\abs{m_{ref}}\abs{m}}.
\end{equation}
We define the polar angles corresponding to the two extremes of a given band as $\theta_{min}$ and $\theta_{max}$ (inset of Fig.~\ref{spheres}). The
width of each band is then ${\Delta}{\theta}={\theta}_{max}-{\theta}_{min}$, and its position defined by $\theta_{min}$. Now we can calculate the
bands' widths and positions by sweeping, for each $\abs{m}$, the values of $\omega_0$ corresponding to ground states (Fig.~\ref{spheres}).

For both systems under study, we find that as $\abs{m}$ increases from $\abs{m_{ref}}$, both ${\theta}_{max}$ and ${\theta}_{min}$ increase. This
has the effect of the bands moving from the north pole to the south pole as we move away from the reference. Additionally, we find that the
bandwidth ${\Delta}{\theta}$ decreases as $\abs{m}$ increases [sketched in Fig.~\ref{spheres}(a), left]. As $\abs{m}$ decreases from $\abs{m_{ref}}$,
we again find that both ${\theta}_{max}$ and ${\theta}_{min}$ increase, with the bands moving from the north pole to the south pole. However, this
time, as $\abs{m}$ decreases, ${\Delta}{\theta}$ increases, meaning that the bands get wider as we move away from the reference [sketched in
Fig.~\ref{spheres}(a), right].

Quantitative results for the ISI system are shown in Fig.~\ref{spheres}(b). We obtain similar results for Hooke's atom (not shown). The band on the
surface of each sphere indicates where all ground state paramagnetic current densities lie within that sphere. In contrast with particle densities or
wavefunctions, we find that, at least for the systems at hand, ground state currents populate a well-defined, limited region of each sphere, whose
size and position display monotonic behavior with respect to the quantum number $m$. This regular behavior is not at all expected, as the CDFT-HK
theorem does not guarantee monotonicity in metric space, and not even that the mapping of $D_{\psi}$ to $D_{\mbf{j}_{p_{\perp}}}$ is single valued.
In the CDFT-HK theorem ground state wavefunctions are uniquely determined only by particle and paramagnetic current densities \emph{together}. In
this sense we can look at the panels in Fig.~\ref{results} as projections on the axis planes of a 3-dimensional
$D_{\psi}D_{\rho}D_{\mbf{j}_{p_{\perp}}}$ relation. The complexity of the mapping due to the application of a magnetic field -- the changes in
quantum number $m$ -- is fully captured by $D_{\mbf{j}_{p_{\perp}}}$ only, as this is related to the relevant conservation law. However the
mapping from $D_{\rho}$ to $D_{\psi}$ inherits the ``band structure,'' showing that the two mappings $D_{\mbf{j}_{p_{\perp}}}$ to $D_{\psi}$ and
$D_{\rho}$ to $D_{\psi}$ are not independent.

\section{Conclusion} \label{conclusion}

In conclusion we showed that conservation laws induce related metric spaces with an ``onion-shell'' geometry and that they may induce a 
``band structure'' in ground state metric spaces, a signature of the enhanced constraints due to the system conservation laws on the relation
between wavefunctions and the relevant physical quantities.

The method proposed may help with understanding extended HK theorems, such as, in the case at hand, the CDFT-HK theorem. In this respect we find
that in metric spaces and for the systems considered, the relevant mappings display distinctive signatures, including (piecewise) linearity at short
and medium distances, the mapping between ground state $\psi$ and $\mbf{j}_{p}$ resembling the one between $\rho$ and $\mbf{j}_{p}$, and the mapping
between ground state $\psi$ and $\mbf{j}_{p}$ showing different trajectories for increasing or decreasing Hamiltonian parameters, in contrast with
the mapping between $\psi$ and $\rho$. Features like this could be used to build or test (single-particle) approximate solutions to many-body
problems, e.g., within DFT schemes.

Our results show that using conservation laws to derive metrics makes these metrics a powerful tool to study many-body systems governed by integral
conservation laws.

\begin{acknowledgments}
 We thank M. Taut, K. Capelle, and C. Verdozzi for helpful discussions. P.~M.~S. acknowledges EPSRC for financial support. I.~D. and P.~M.~S.
 gratefully acknowledge support from a University of York - FAPESP combined grant.
\end{acknowledgments}

\bibliography{References}
\end{document}